\documentclass[11pt]{article}
\usepackage{latexsym,times}
\usepackage{amsmath}
\usepackage{amsxtra}
\usepackage{amscd}
\usepackage{amsthm}
\usepackage{amsfonts}
\usepackage{amssymb}

\usepackage{esint}

\usepackage{enumitem}
\usepackage{float}

\usepackage{hyperref}
\usepackage[square,sort&compress,comma,numbers]{natbib}

\makeatletter

\@addtoreset{equation}{section} \makeatother

\input amssym.def
\input amssym.tex

\newcommand{\half}{{{\textstyle\frac{1}{2}}}}
\newcommand{\quarter}{{{\textstyle\frac{1}{4}}}}
\newcommand{\be}{\begin{equation}}
\newcommand{\ee}{\end{equation} }
\newcommand{\beqa}{\begin{eqnarray} }
\newcommand{\eeqa}{\end{eqnarray} }
\newcommand{\ba}{\begin{array}}
\newcommand{\ea}{\end{array}}
\newcommand{\bpm}{\begin{pmatrix}}
\newcommand{\epm}{\end{pmatrix}}

\newcommand{\SO}{\mathbf{SO}}

\newcommand{\ODD}{\mathbf{O}(D,D)}

\newcommand\tr{{\rm tr\,}}

\newcommand\rd{{\rm d}}

\newcommand\rdA{D}

\newcommand\cA{{\cal A}}

\newcommand\cC{{\cal C}}

\newcommand\cE{{\cal E}}

\newcommand\cH{{\cal H}}

\newcommand\cJ{{\cal J}}

\newcommand\cL{{\cal L}}

\newcommand\cN{{\cal N}}
\newcommand\cO{{\cal O}}
\newcommand\cP{{\cal P}}

\newcommand\cS{{\cal S}}

\newcommand\cV{{\cal V}}

\newcommand\bcP{{\bar{\cP}}}

\newcommand\hcL{{\hat{\cal L}}}

\newcommand\dis{\displaystyle}

\def\tx{\tilde{x}}
\def\ty{\tilde{y}}

\def\tY{\tilde{Y}}

\def\brF{\bar{F}}

\def\brL{\bar{L}}

\def\brP{{\bar{P}}}

\def\Tw{{T_{\omega}}}

\newcommand{\DO}{\mathbf{\nabla}}

\newcommand{\na}{{\nabla}}

\newcommand{\dsR}{{\mathbb R}}

\begin{document}
\begin{titlepage}
\title{\vskip -100pt
\vskip 2cm Covariant action for a string  in  {\textit{doubled yet gauged}} spacetime\\}
\author{\sc{Kanghoon Lee${}^{\ast}$  ~and~ Jeong-Hyuck Park${}^{\dagger}$}}
\date{}
\maketitle \vspace{-1.0cm}
\begin{center}
${}^{\ast}$Blackett Laboratory, Imperial College, London SW7 2AZ,  England\\
Center for Quantum Spacetime, Sogang University,  Seoul 121-742, Korea\\
~\\
${}^{\dagger}$Department of Applied Mathematics and Theoretical Physics, Cambridge, CB3 0WA, England\\
Department of Physics, Sogang University,  Seoul 121-742, Korea\footnote{Sabbatical leave of absence.}\\
~\\
\texttt{kanghoon.lee@imperial.ac.uk\quad \quad park@sogang.ac.kr}
~{}\\
~~~\\~\\
\end{center}
\begin{abstract}
\vskip0.1cm  \noindent The section condition in double field theory has been   shown   to imply    that  a physical point should be   one-to-one identified with a    gauge orbit in the  doubled coordinate space. Here we  show     the converse is also true,  and  continue to   explore     the idea of  \textit{spacetime being  doubled yet gauged}. Introducing    an appropriate   gauge connection,  we construct     a   string action,   with  an arbitrary generalized metric,    which   is completely   covariant  with respect to  the coordinate gauge symmetry,   generalized diffeomorphisms, world-sheet diffeomorphisms, world-sheet Weyl symmetry   and  $\ODD$ T-duality. A   topological term   previously     proposed in the literature   naturally arises    and  a  self-duality condition  follows  from the equations of motion. Further,  the action may  couple to   a T-dual  background  where the Riemannian metric becomes    everywhere singular.    
\end{abstract}

{\small
\begin{flushleft}
~~
\end{flushleft}}
\thispagestyle{empty}
\end{titlepage}
\newpage
\tableofcontents 

\section{Introduction}
In order to realize  $\ODD$ T-duality as a manifest symmetry~\cite{Duff:1989tf,Tseytlin:1990nb,Tseytlin:1990va,Rocek:1991ps,Giveon:1991jj,Hull:2004in,Hull:2006qs,Hull:2006va},  Double Field Theory (DFT)~\cite{Siegel:1993xq,Siegel:1993th,Hull:2009mi,Hull:2009zb,Hohm:2010jy,Hohm:2010pp}  doubles    the spacetime dimension,  from $D$ to ${D+D}$,  with  doubled coordinates,   $x^{A}=(\ty_{\mu},y^{\nu})$, of which the first  and the last   correspond to   the `winding' and the  `ordinary' coordinates respectively.  However, DFT is not truly doubled since all the fields ---including any  local symmetry parameters--- are subject to the section condition: the $\ODD$ invariant d'Alembertian operator must be  trivial, acting on arbitrary fields,  
\be
\partial_{A}\partial^{A}\Phi(x)=0\,,
\label{wcon}
\ee
as well as their products, or equivalently 
\be
\partial_{A}\Phi_{1}(x)\partial^{A}\Phi_{2}(x)=0\,.
\label{stcon}
\ee
While the section condition might appear somewhat odd or even invidious  from the conventional Riemannian point of view, it is readily satisfied when all the fields are, up to $\ODD$ rotations,   independent of the dual winding coordinates, \textit{i.e.~}$\frac{\partial~~}{\partial\ty_{\mu}}\equiv 0$. This kind of explicit `section-fixing'     reduces DFT to \textit{generalized geometry}~\cite{Hitchin:2004ut,Hitchin:2010qz,Gualtieri:2003dx,Pacheco:2008ps,Grana:2008yw,Coimbra:2011nw,Coimbra:2011ky,Coimbra:2012yy,Coimbra:2012af,Koerber:2010bx} where the spacetime  is not enlarged   and the duality is less manifest. \\

\noindent Much progress  has been made in recent years    based on  the notion of    doubled  spacetime subject to the  section condition~\cite{Jeon:2010rw,Hohm:2010xe,Jeon:2011kp,Jeon:2011cn,Thompson:2011uw,Copland:2011yh,Hohm:2011dv,Albertsson:2011ux,Jeon:2011vx,Geissbuhler:2011mx,Aldazabal:2011nj,Berman:2011kg,Copland:2011wx,Jeon:2011sq,Copland:2012zz,Grana:2012rr,Dibitetto:2012rk,Copland:2012ra,Jeon:2012kd,Hohm:2012gk,Nibbelink:2012jb,Jeon:2012hp,Hohm:2012mf,Aldazabal:2013mya,Berman:2013uda,Geissbuhler:2013uka,Park:2013mpa,Aldazabal:2013sca,Berman:2013cli,Berman:2013eva,Hohm:2013jaa,Andriot:2013xca,Godazgar:2013bja,Hohm:2013nja,Wu:2013sha},   including 
the state of the art reviews~\cite{Aldazabal:2013sca,Berman:2013eva} and    the construction of  ${\cN=2}$ ${D=10}$  maximally supersymmetric double field theory~\cite{Jeon:2012hp} as the unification of   type IIA and IIB supergravities.\footnote{\textit{c.f.} {\url{http://strings2013.sogang.ac.kr//main/?skin=video_GS_2.htm}}}  Analogous   parallel developments   on   U-duality are also available~\cite{Berman:2010is,Berman:2011pe,Berman:2011cg,Berman:2011jh,Malek:2012pw,Berman:2012uy,Berman:2012vc,Musaev:2013rq,Malek:2013sp,Park:2013gaj,Cederwall:2013naa,Cederwall:2013oaa,Godazgar:2013rja,Hohm:2013jma}\footnote{\textit{c.f.}     
{\url{http://strings2013.sogang.ac.kr//main/?skin=video_27_5.htm}}} which may all be   incorporated into the \textit{grand scheme of}  $E_{11}$~\cite{West:2001as,West:2010ev,Rocen:2010bk,West:2011mm}.  \\

\noindent The ${(D+D)}$-dimensional doubled  spacetime is  far from being   an ordinary Riemannian manifold,  since  it  postulates  the existence of a  globally well-defined $\ODD$ invariant constant metric,
\be
\cJ_{AB}=\left(\ba{cc}0&1\\1&0\ea\right)\,,
\label{ODDmetric}
\ee
which serves to raise and lower  the doubled spacetime indices.\footnote{For example, $\partial^{A}=\cJ^{-1AB}\partial_{B}$ as done in (\ref{wcon}) and (\ref{stcon}).}  Further,  the diffeomorphism symmetry   is  generated  not by the ordinary Lie derivative but  by a  generalized one,
\be
\hcL_{V}T_{A_{1}\cdots A_{n}}:=V^{B}\partial_{B}T_{A_{1}\cdots A_{n}}+\omega\partial_{B}V^{B}T_{A_{1}\cdots A_{n}}+\sum_{i=1}^{n}(\partial_{A_{i}}V_{B}-\partial_{B}V_{A_{i}})T_{A_{1}\cdots A_{i-1}}{}^{B}{}_{A_{i+1}\cdots  A_{n}}\,,
\label{tcL}
\ee
where   $\omega$ is the weight of the DFT-tensor,  $T_{A_{1}\cdots A_{n}}(x)$,   and  $V^{A}(x)$ is an  {infinitesimal diffeomorphism  parameter}, as a DFT vector field which must also obey   the section condition, 
\be
\ba{ll}
\partial_{A}\partial^{A}V^{B}(x)=0\,,\quad&\quad \partial_{A}V^{B}(x)\partial^{A}\Phi(x)=0\,.
\ea
\label{secconV}
\ee
The generalized Lie derivative of the $\ODD$ metric vanishes  for consistency, 
\be
\hcL_{V}\cJ_{AB}=0\,.
\label{tcLJ}
\ee
~\\

\noindent As pointed out  in \cite{Park:2013mpa},  the section condition implies that  the coordinates  in doubled spacetime do not    represent the   physical points in an injective manner. Rather,   a physical point should  be     one-to-one identified   with   a `gauge orbit' in the coordinate space,  \textit{i.e.}        an  \textit{equivalence relation} holds for the doubled coordinates:
\be
x^{A}~\sim~ x^{A}+\phi\partial^{A}\varphi\,,
\label{equiv0}
\ee
where $\phi$ and $\varphi$ are two  arbitrary functions in DFT. While we review the  explicit realization of this equivalence below, \textit{c.f.}~\eqref{TensorCGS},  it implies that  \underline{\textbf{\textit{spacetime is doubled yet gauged.}}}    
~The diffeomorphism symmetry then means  an invariance under arbitrary reparametrizations of the gauge orbits.  
Henceforth, we call the equivalence relation on coordinates~(\ref{equiv0}), `coordinate gauge symmetry' so that the quotient/equivalence classes   form a space diffeomorphic to the section. It follows that  a similar  equivalence relation  holds     for  the infinitesimal diffeomorphism parameters, 
\be
V^{A}~~\sim~~V^{A}+\phi\partial^{A}\varphi\,,
\label{cgs2}
\ee
and consequently there is  more than one finite  tensorial  diffeomorphic transformation rule,  while the simplest  choice seems to be the one found  in \cite{Hohm:2012gk} which we shall   recall in (\ref{GD}).  In fact, it was   the coordinate gauge symmetry that  played a crucial role to resolve a puzzle left in \cite{Hohm:2012gk} where the simple tensorial  diffeomorphic transformation rule found therein did not  coincide  with  the exponentiation of the generalized Lie derivative~(\ref{tcL}).  Nevertheless,   they are equivalent  up to the coordinate gauge symmetry~\cite{Park:2013mpa}.\\

\noindent It is the purpose of this work  to further explore   the geometric significance of the coordinate gauge symmetry~(\ref{equiv0}), in particular by considering a string which propagates in a doubled yet gauged spacetime. Compared to preceding  works on sigma models in doubled space~\cite{Duff:1989tf,Tseytlin:1990nb,Tseytlin:1990va,Hull:2006qs,Hull:2006va,Copland:2011wx,Nibbelink:2012jb}, the novelties  of the action to be constructed in this paper,  (\ref{Lagrangian}),   are  as follows. \textit{i)}  As in DFT, \textit{a priori}  no  isometry nor torus structures   are assumed.  For an arbitrary given  background, our action is fully covariant under world-sheet diffeomorphism, world-sheet  Weyl symmetry, target spacetime generalized diffeomorphisms, $\ODD$ T-duality and the coordinate gauge symmetry. \textit{ii)} The full spacetime dimensions are doubled,  yet  they are gauged. \textit{iii)} The self-duality relation follows from the equations of motion of the auxiliary gauge fields, without breaking any symmetry.    \textit{iv)} The action  may also describe a string which propagates in  a novel class of non-Riemannian geometries. \newpage

The rest of the paper  is organized as follows.
\begin{itemize}
\item In section~\ref{secCGS},  we prove that the coordinate gauge symmetry~(\ref{equiv0}) implies the section condition, both  (\ref{wcon}) and  (\ref{stcon}). Hence, with the result of \cite{Park:2013mpa}, they are in fact equivalent. This motivates us to propose to take    the `coordinate gauge symmetry' as the  geometrical  first  principle for the doubled spacetime formalism.  
\item In section~\ref{secGAUGING}, we explicitly introduce a gauge connection for the coordinate gauge symmetry and define \textit{gauged differential one-forms for the doubled coordinates}, $\rdA x^{M}=\rd x^{M}-\cA^{M}$. We demonstrate  their covariant properties under  generalized diffeomorphisms and the coordinate gauge symmetry.
\item In section~\ref{secString}, in terms of the gauged differential one-forms, we construct an  \textit{action for a string     in the doubled yet gauged spacetime}, which is completely covariant  with respect to   the coordinate gauge symmetry,   generalized diffeomorphisms, $\ODD$ T-duality, world-sheet diffeomorphisms and  world-sheet Weyl symmetry.
\item In section~\ref{secReduction},  we discuss reductions to undoubled formalisms. We point out that  there are generically two classes of reductions depending on the generalized metric. For the \textit{non-degenerate} case, the generalized metric can as usual be  parametrized by a Riemannian metric and a Kalb-Ramond $B$-field. The reduction then naturally  recovers the  doubled sigma model proposed in \cite{Hull:2006va} including a `topological term'~\cite{Giveon:1991jj} and a `self-duality' relation~\cite{Tseytlin:1990nb,Tseytlin:1990va}. On the other hand, the  \textit{degenerate}  case  deals with a background  where  the Riemannian metric, if interpreted in that way,  is everywhere singular.  As an example, we obtain such a singular  background   by T-dualizing a fundamental string geometry~\cite{Dabholkar:1990yf}.
\item We conclude with a \textbf{\textit{summary}} and comments \textbf{\textit{in section~\ref{secCON}}}. 
\item Appendices contain a brief  review of covariant derivatives in DFT  and some useful formulae.
\end{itemize}

While the idea of spacetime being gauged might sound strange at first, since the spacetime coordinates are dynamic fields on world-sheet, the coordinate gauge symmetry will be realized as just one of the  gauge symmetries of the constructed action along  with others.

\section{\textit{Doubled yet gauged} spacetime\label{secCGS}}
\textit{Coordinate gauge symmetry}  means an equivalence relation on the coordinates  of the doubled spacetime which is generated by a \textit{derivative-index-valued} shift~\cite{Park:2013mpa},
\be
x^{A}~\sim~ x^{A}+\phi\partial^{A}\varphi\,,
\label{equivCGS}
\ee
where $\phi$ and $\varphi$ are two  arbitrary functions in DFT. The \textit{coordinate gauge symmetry} is additive,  being Abelian in nature, 
\be
\ba{lll}
x^{A}~\sim~ x^{A}+\phi\partial^{A}\varphi\,,~\quad x^{A}~\sim~ x^{A}+\phi^{\prime}\partial^{A}\varphi^{\prime}\quad&\Longrightarrow&\quad x^{A}~\sim~ x^{A}+\phi\partial^{A}\varphi+\phi^{\prime}\partial^{A}\varphi^{\prime}\,.
\ea
\ee
All the functions of DFT, \textit{i.e.~}field variables,   local symmetry parameters  (including $\phi$ and $\varphi$ in \eqref{equivCGS}) and  their arbitrary derivative-descendants,  are then  by definition  required to be invariant under the derivative-index-valued shift,
\be
\ba{ll}
\partial_{A_{1}}\partial_{A_{2}}\cdots\partial_{A_{n}}\Phi(x+\Delta)=\partial_{A_{1}}\partial_{A_{2}}\cdots\partial_{A_{n}}\Phi(x)\,,\quad&\quad \Delta^{A}=\phi\partial^{A}\varphi\,,
\ea
\label{TensorCGS}
\ee
where $n=0,1,2,\cdots$. \\

\noindent For a generic (local) shift of the coordinates, with an arbitrary real parameter, $s\in\dsR$, 
\be
\ba{lll}
x^{A}\quad&\longrightarrow&\quad x^{A}+s\Delta^{A}(x)\,,
\ea
\ee 
a standard power expansion reads
\be
T_{A_{1}\cdots A_{m}}(x+s\Delta)=T_{A_{1}\cdots A_{m}}(x)+{\sum_{l=1}^{\infty}}\,\frac{s^{l}}{l!}\,\Delta^{B_{1}}\Delta^{B_{2}}\cdots\Delta^{B_{l}}\partial_{B_{1}}\partial_{B_{2}}\cdots\partial_{B_{l}}T_{A_{1}\cdots A_{m}}(x)\,.
\label{expDelta}
\ee
From the consideration of \textit{e.g.}~putting   $T_{A_{1}\cdots A_{m}}=\partial_{A_{1}}\cdots\partial_{A_{m}}\Phi_{1}$ and   $\Delta^{A}=k^{B_{1}}\cdots k^{B_{n}}\partial_{B_{1}}\cdots\partial_{B_{n}}\partial^{A}\Phi_{2}$ with an arbitrary constant vector, $k^{B}$, it follows immediately   that  the coordinate gauge symmetry~(\ref{equivCGS}) implies the section condition  which is quadratic in functions~(\ref{stcon}),
\be
\ba{ll}
\left(\partial_{A_{1}}\partial_{A_{2}}\cdots\partial_{A_{m}}\partial^{C}\Phi_{1}\right)
\left(\partial_{B_{1}}\partial_{B_{2}}\cdots\partial_{B_{n}}\partial_{C}\Phi_{2}\right)=0\,,\quad&\quad
m,n\geq 0\,.
\ea
\ee
Further, as we show shortly, a particular  case of this result ($\Phi_{1}=\Phi_{2}$ and $m=n=1$)  leads to  the other  section condition, or the linear ``weak" constraint~(\ref{wcon}). Hence, the coordinate gauge symmetry~(\ref{equivCGS}) implies the section condition, both (\ref{wcon}) and (\ref{stcon}).   Since the converse is also true from (\ref{TensorCGS}) and  (\ref{expDelta})~\cite{Park:2013mpa}, we conclude the following.  \\

\begin{tabular}{|c|}
\hline
\textit{~The coordinate gauge symmetry~(\ref{equivCGS}) is equivalent to the section condition, both (\ref{wcon}) and (\ref{stcon}),\,~}\\
{}$x^{A}~\sim~ x^{A}+\phi\partial^{A}\varphi~~~~~~\Longleftrightarrow~~~~~~
\partial_{A}\partial^{A}\Phi(x)=0~~~~\mbox{\&}~~~~
\partial_{A}\Phi_{1}(x)\partial^{A}\Phi_{2}(x)=0\,,$\\
\textit{and serves as  a geometric first principle for the doubled spacetime formalism. }~\quad\quad\quad\quad\quad\quad\quad\\ 
\hline
\end{tabular}

~\\
~\\
\indent{\bf{\textit{Theorem.~}}} \underline{If $\,\partial_{A}\partial_{B}\Phi\partial^{A}\partial_{C}\Phi=0\,$ then $\,\partial_{A}\partial^{A}\Phi=0$.}\\
\indent {\bf{\textit{Proof.~}}}    The given assumption implies  the nilpotent property of a ${(D+D)}\times {(D+D)}$ square matrix, 
\be
\ba{ll}
M_{A}{}^{B}=\partial_{A}\partial^{B}\Phi\,,\quad&\quad M^{2}=0\,.
\ea
\ee
Hence, with an arbitrary real parameter, $s\in\dsR$, we have
\be
\det(1+sM)=e^{\tr\ln(1+sM)}=e^{s\,\tr M}\,,
\ee
or
\be
\det(\delta_{A}{}^{B}+s\partial_{A}\partial^{B}\Phi)=e^{s\partial_{A}\partial^{A}\Phi}=1+\sum_{n=1}^{\infty}\frac{s^{n}}{n!}(\partial_{A}\partial^{A}\Phi)^{n}\,.
\label{det1s}
\ee
Since  the determinant  is a finite polynomial in the variable, $s$, while the exponential has an \textit{a priori} infinite power series expansion,  it is clear that  each higher order term of the latter   must vanish, or
\be
\partial_{A}\partial^{A}\Phi=0\,.
\label{LapF}
\ee
In particular,  the determinant is one, being $s$-independent.  This  completes  our proof.  \\

{\bf{\textit{Comments.~}}}  \textit{(i)} An alternative proof may be established by considering  the  `{Jordan normal form}' of the square  matrix, $M$. The nilpotent property of the  matrix implies that all the diagonal elements of its  Jordan normal form  are zero,  and hence the matrix is traceless~(\ref{LapF}).\\
\indent \textit{(ii)} With the relation,
\be
0=\int_{\dsR^{D+D}} \partial_{A}\partial_{B}\Phi\partial^{A}\partial^{B}\Phi=
\int_{\dsR^{D+D}} \partial_{A}(\partial_{B}\Phi\partial^{A}\partial^{B}\Phi-\partial^{A}\Phi\partial_{B}\partial^{B}\Phi)+(\partial_{A}\partial^{A}\Phi)^{2}\,,
\ee
if  we assume  ``sufficiently fast fall off behavior at infinity" in order to ignore  the total derivative or the surface integral,  with the positive definite property of $(\partial_{A}\partial^{A}\Phi)^{2}$ we might argue for (\ref{LapF}). However, this assumption appears too strong to be realized in double field theory where explicitly the fields do not depend on the dual  winding coordinates. Thus, it is desirable to establish a direct \textit{proof}, as presented above,  which holds irrespective of the  boundary  conditions.

 \textit{(iii)} It is worthwhile to note that, unlike $\partial_{A}\partial_{B}\Phi\partial^{A}\partial_{C}\Phi=0$, an alternative condition, $\,\partial_{A}\Phi\partial^{A} \Phi=0$, does not necessarily imply the weak constraint, $\partial_{A}\partial^{A}\Phi=0$, except in dimension ${D=1}$, since a  counterexample   exists for $D\geq 2$, 
\be
\Phi^{\prime}(x^{1},\cdots,x^{D},\tx_{1},\cdots,\tx_{D})=\exp\left[2\sqrt{x^{1}\tx_{1}}+\sum_{\mu=2}^{D}\frac{\,x^{\mu}-\tx_{\mu}\,}{\sqrt{D-1}}\right]\,,
\ee
which satisfies 
\be
\ba{ll}
\partial_{A}\Phi^{\prime}\partial^{A} \Phi^{\prime}=0\,,\quad&\quad
\partial_{A}\partial^{A}\Phi^{\prime}=\frac{1}{2\sqrt{x^{1}\tx_{1}}}\Phi^{\prime}\neq 0\,.
\ea
\ee
~\\

\section{Gauge connection for the coordinate gauge symmetry \label{secGAUGING}}
We recall   two  finite   local symmetries  of  double field theory.
\begin{itemize}
\item  \textit{Generalized diffeomorphism}, $\,x^{M}\rightarrow x^{\prime M}$, in the `passive' form~\cite{Hohm:2012gk,Park:2013mpa},
\be
\ba{lll}
T_{A_{1}A_{2}\cdots A_{n}}(x)~~&\longrightarrow&~~T^{\prime}_{A_{1}A_{2}\cdots A_{n}}(x^{\prime})=\left(\det L\right)^{-\omega}\brF_{A_{1}}{}^{B_{1}}\brF_{A_{2}}{}^{B_{2}}\cdots \brF_{A_{n}}{}^{B_{n}}
T_{B_{1}B_{2}\cdots B_{n}}(x)\,,
\ea
\label{GD}
\ee
where 
\be
\ba{ll}
L_{A}{}^{B}=\partial_{A}x^{\prime B}\,,~~~~&~~~~\brL=\cJ L^{t}\cJ^{-1}\,,\\
F=\half\left(L\brL^{-1}+\brL^{-1}L\right)\,,~~~~&~~~~
\brF=\cJ F^{t}\cJ^{-1}=\half\left(L^{-1}\brL+\brL L^{-1}\right)\,.
\ea
\label{LbrLF}
\ee
In particular,  $F$ can be shown to be  an $\ODD$ element,
\be
\ba{ll}
F\brF=1\,,~~~~&~~~~ F\in\ODD\,,
\ea
\ee
which  agrees   with (\ref{tcLJ}). It should be also noted that  it is not the transformed coordinates, $x^{\prime M}$,  but the difference, $x^{\prime M}-x^{M}$,   that  satisfies the section condition, (\ref{wcon}), (\ref{stcon}),\footnote{For example,  $\,\partial_{A}x^{M}\partial^{A}\Phi=\partial^{M}\Phi\neq0$.} such that, \textit{e.g.}
\be
\partial^{A}\Phi(L-1)_{A}{}^{B}=0\,.
\ee
\item \textit{Coordinate gauge symmetry}~(\ref{equivCGS})~\cite{Park:2013mpa},
\be
\ba{ll}
x^{A}~\,\sim\,~ x_{s}^{M}=e^{s\cV{\cdot\partial}}x^{M}=x^{M}+s\cV^{M}(x)\,,\quad&\quad
T_{A_{1}A_{2}\cdots A_{n}}(x)=T_{A_{1}A_{2}\cdots A_{n}}(x_{s})\,,
\ea
\label{fCGSx}
\ee
which is generated  by  a \textit{derivative-index-valued} vector, $\cV^{A}$, 
\be
\cV^{M}=\phi\partial^{M}\varphi\,,
\label{divV}
\ee
satisfying the section condition~(\ref{stcon}),
\be
\cV^{M}\partial_{M}\Phi=0\,.
\ee
\end{itemize}
The main difference of the above two local symmetries     is  formally whether the spacetime indices of  the tensors are  supposed to be rotated or not.\footnote{Of course,  the transformation, $x^{M}\,\rightarrow\,x_{s}^{M}=x^{M}+s\cV^{M}(x)$, can be taken as a sort of  generalized diffeomorphism  which can be then shown  to reduce to the $B$-field gauge symmetry only, without involving any   Riemannian  diffeomorphism~\cite{Park:2013mpa}.} \\

\noindent We now introduce a gauge connection,\footnote{A closely related earlier work is \cite{Rocek:1991ps} where `gauging'  chiral currents  on a world-sheet was discussed, see also \cite{Giveon:1991jj,Hull:2006va}.  } $\cA^{M}$, and define  \textit{gauged differential one-forms for the doubled coordinates,}
\be
\rdA x^{M}:=\rd x^{M}-\cA^{M}\,.
\label{GDO}
\ee
We require the connection to be a \textit{derivative-index valued} one-form,  satisfying
\be
\ba{ll}
\cA^{M}\partial_{M}\Phi=0\,,\quad&\quad \cA^{M}(L-1)_{M}{}^{N}=0\,.
\ea
\label{divA}
\ee
~\\
Under the  finite generalized diffeomorphism~(\ref{GD}), the gauge connection must   transform as
\be
\ba{lll}
\cA^{M}\quad&\longrightarrow&\quad \cA^{\prime M}=\cA^{N}F_{N}{}^{M}+\rd x^{N}(L-F)_{N}{}^{M}\,,
\ea
\label{GDA}
\ee
such that the gauged differential one-forms are \textit{covariant,}
\be
\ba{lll}
\rdA x^{M}\quad&\longrightarrow&\quad \rdA^{\prime}x^{\prime M}=\rdA x^{N}F_{N}{}^{M}\,.
\ea
\ee
Furthermore, thanks to the following identities  which derive   from the section condition~\cite{Park:2013mpa} (see also \cite{Hohm:2012gk}),
\be
\ba{l}
(1-\brL^{-1})L=1-\brL^{-1}\,,\\
L-F=\half(L+1)(1-\brL^{-1})L=\half(L+1)(1-\brL^{-1})\,,\\
F\brL-1=\half(L-1)(1+\brL)\,,\\
\partial_{A}\Phi=L_{A}{}^{B}\partial_{B}^{\prime}\Phi=F_{A}{}^{B}\partial_{B}^{\prime}\Phi\,,\\
(1-\brL)_{A}{}^{B}\partial_{B}\Phi=(1-\brL^{-1})_{A}{}^{B}\partial_{B}\Phi=0\,,
\ea
\label{LmF}
\ee
the gauge  connection   remains still  \textit{derivative-index-valued}  after the transformation~(\ref{GDA}).  In other words,   the derivative-index-valuedness   of the connection   is preserved under  generalized diffeomorphisms,
\be
\dis{\cA^{\prime M}\partial^{\prime}_{M}\Phi=
\cA^{M}\partial_{M}\Phi=0\,.}
\ee
With  the (passive) tensorial  transformation rule~(\ref{GD}),
\be
\ba{lll}
\cH_{MN}\quad&\longrightarrow&\quad \cH_{MN}^{\prime}=\brF_{M}{}^{K}\brF_{N}{}^{L}\cH_{KL}\,,
\ea
\ee
we can  now  ensure  the invariance of the generalized metric under generalized diffeomorphisms, 
\be
\cH_{MN}\rdA x^{M}\otimes\rdA x^{N}=\cH^{\prime}_{MN}\rdA^{\prime}x^{\prime M}\otimes\rdA^{\prime}x^{\prime N}\,.
\ee
~\\
On the other hand,  under the coordinate gauge symmetry~(\ref{fCGSx}),  the gauge connection transforms  precisely the same way as the one-form, $\rd x^{M}$,  
\be
\ba{ll}
\rd x^{M}\quad\longrightarrow\quad\rd x_{s}^{M}=\rd x^{M}+s\rd(\phi\partial^{M}\varphi)\,,\quad&\quad \cA^{M}\quad\longrightarrow\quad\cA_{s}^{M}=\cA^{M}+s\rd(\phi\partial^{M}\varphi)\,,
\ea
\ee
such that it preserves  the derivative-index-valuedness.  The gauged differential one-forms are then simply   \textit{invariant,}
\be
\rdA x^{M}=\rdA_{s} x_{s}^{M}\,.
\label{invcgs}
\ee
~\\

\section{Completely covariant  string action \label{secString}}
We pull back the gauged differential one-forms~(\ref{GDO}) to   a string world-sheet with coordinates, $\sigma^{i}$, $i=0,1$, and  promote  the doubled target spacetime  coordinates and the gauge connection  to the world-sheet fields, $X^{M}(\sigma)$ and $\cA_{i}^{M}(\sigma)$. That is, $\sigma^{i}$  denotes a coordinate on the world-sheet, $\Sigma$,  and $X^M$ are the components of $X:\Sigma \to {\mathbf{R}}^{D+D}$ so that
\be
\rdA X^{M}=\rd\sigma^{i}\rdA_{i}X^{M}=\rd\sigma^{i}(\partial_{i}X^{M}-\cA_{i}^{M})\,.
\ee
The string  action  we construct in this work is then
\be
\ba{ll}
\cS={\textstyle{\frac{1}{4\pi\alpha^{\prime}}}}{\dis{\int_{\Sigma}}}\rd^{2}\sigma~\cL\,,\quad&\quad
\cL=-\half\sqrt{-h}h^{ij}\rdA_{i}X^{M}\rdA_{j}X^{N}\cH_{MN}(X)-\epsilon^{ij}\rdA_{i}X^{M}\cA_{jM}\,.
\ea
\label{Lagrangian}
\ee
Here $h_{ij}$ corresponds to the usual auxiliary world-sheet  metric which can raise or lower the world-sheet coordinate indices, $i,j$. The action  describes a string propagating  in  a doubled yet gauged spacetime with an arbitrarily given  generalized metric, $\cH_{AB}(X)$.  Apart from the section condition, the generalized metric   only needs to satisfy  the  two defining properties, 
\be
\ba{ll}
\cH_{AB}=\cH_{BA}\,,\quad&\quad\cH_{A}{}^{B}\cH_{B}{}^{C}=\delta_{A}{}^{C}\,.
\ea
\label{defcH}
\ee
Otherwise it is quite arbitrary. The string tension  in the doubled spacetime   is  halved, \textit{i.e.~}$({4\pi\alpha^{\prime}})^{-1}$ instead of $({2\pi\alpha^{\prime}})^{-1}$ as  stressed     in  \cite{Hull:2006va}.  It may recover the standard value, $({2\pi\alpha^{\prime}})^{-1}$, after  reduction to an undoubled  formalism, \textit{c.f.~}(\ref{tensionLprime}). In fact, as we see in the next section,  \textit{c.f.~}(\ref{cCcA}), at least for  ``non-degenerate" cases of the generalized metric,    the  above doubled string action precisely reduces  to the standard undoubled  string action with the right number of degrees of freedom.  While we reserve section~\ref{secReduction} for the exposition of the reductions to   undoubled formalisms, in the remaining of the current section we focus on  the  covariant properties of the doubled  action. 

The Lagrangian is manifestly  symmetric with respect to  the $\ODD$ T-duality and  the  world-sheet diffeomorphisms.  Furthermore, from (\ref{divA}), (\ref{LmF}) and (\ref{invcgs}),    up to  total  derivatives  it  is invariant  under  the coordinate gauge symmetry  as
\be
\ba{ll}
\epsilon^{ij}\rdA_{si}X^{M}_{s}\cA_{sjM}&=
\epsilon^{ij}\rdA_{i}X^{M}\left[\cA_{jM}
+s\partial_{j}X^{N}\partial_{N}(\phi\partial_{M}\varphi)\right]\\
{}&=\epsilon^{ij}\rdA_{i}X^{M}\cA_{jM}
+s\epsilon^{ij}\partial_{i}X^{M}\partial_{j}X^{N}\partial_{N}(\phi\partial_{M}\varphi)\\
{}&=\epsilon^{ij}\rdA_{i}X^{M}\cA_{jM}
+\partial_{j}\left[
s\epsilon^{ij}\partial_{i}X^{M}(\phi\partial_{M}\varphi)\right]\\
{}&=\epsilon^{ij}\rdA_{i}X^{M}\cA_{jM}
-\partial_{i}\left(
s\epsilon^{ij}\phi\partial_{j}\varphi\right)\,,
\ea
\ee
and also invariant under the generalized diffeomorphisms  as
\be
\ba{ll}
\epsilon^{ij}\rdA^{\prime}_{i}X^{\prime M}\cA^{\prime}_{jM}&=
\epsilon^{ij}\rdA_{i}X^{L}F_{L}{}^{M}\left[\brF_{M}{}^{N}\cA_{jN}+(\brL-\brF)_{M}{}^{N}\partial_{j}X_{N}\right]\\
{}&=\epsilon^{ij}\rdA_{i}X^{M}\cA_{jM}+\epsilon^{ij}\rdA_{i}X^{M}(F\brL-1)_{M}{}^{N}\partial_{j}X_{N}\\
{}&=\epsilon^{ij}\rdA_{i}X^{M}\cA_{jM}+\half\epsilon^{ij}\partial_{i}X^{M}\left[(L-1)(1+\brL)\right]_{M}{}^{N}\partial_{j}X_{N}\\
{}&=\epsilon^{ij}\rdA_{i}X^{M}\cA_{jM}+\half\epsilon^{ij}\partial_{i}X^{M}\partial_{j}X^{N}
(L\cJ-\cJ L^{t}+L\cJ L^{t}-\cJ)_{[MN]}\\
{}&=\epsilon^{ij}\rdA_{i}X^{M}\cA_{jM}+\half\epsilon^{ij}\partial_{i}X^{M}\partial_{j}X^{N}
(L\cJ-\cJ L^{t})_{[MN]}\\
{}&=\epsilon^{ij}\rdA_{i}X^{M}\cA_{jM}+\epsilon^{ij}\partial_{i}X^{M}
L_{M}{}^{N}\partial_{j}X_{N}\\
{}&=\epsilon^{ij}\rdA_{i}X^{M}\cA_{jM}+\epsilon^{ij}\partial_{i}X^{\prime N}\partial_{j}X_{N}\\
{}&=\epsilon^{ij}\rdA_{i}X^{M}\cA_{jM}+\partial_{i}\left(\epsilon^{ij}X^{\prime N}\partial_{j}X_{N}\right)\,.
\ea
\ee
~\\

\noindent  Generically,     under arbitrary variations of all the fields, the Lagrangian transforms as
\be
\ba{ll}
\delta\cL~=\!&-\half\sqrt{-h}\delta h^{ij}\left(\rdA_{i}X^{M}\rdA_{j}X^{N}-\half h_{ij}\rdA_{k}X^{M}\rdA^{k}X^{N}\right)\cH_{MN}\\
{}&+\delta X^{L}\left[\partial_{i}\left(\sqrt{-h}\rdA^{i}X^{M}\cH_{ML}-\epsilon^{ij}\rdA_{j}X_{L}\right)
-\half\sqrt{-h}\rdA_{i}X^{M}\rdA^{i}X^{N}\partial_{L}\cH_{MN}\right]\\
{}&+\delta\cA_{iL}\left[\sqrt{-h}\cH^{L}{}_{M} \rdA^{i}X^{M}+\epsilon^{ij}\partial_{j}X^{L}\right]\\
{}&-\partial_{i}\left[\delta X^{L}\left(\sqrt{-h}\rdA^{i}X^{M}\cH_{LM}+\epsilon^{ij}\cA_{jL}\right)\right]\,,
\ea
\label{forEOM}
\ee
where the second  line can be rewritten  in  an alternative manner, 
\be
\ba{l}
\delta X^{L}\left[\partial_{i}(\sqrt{-h}\rdA^{i}X^{M}\cH_{ML}-\epsilon^{ij}\rdA_{j}X_{L})
-\half\sqrt{-h}\rdA_{i}X^{M}\rdA^{i}X^{N}\partial_{L}\cH_{MN}\right]\\
=\delta X^{L}\left[\cH_{LM}\partial_{i}(\sqrt{-h}\rdA^{i}X^{M})
+\sqrt{-h}\rdA_{i}X^{M}\rdA^{i}X^{N}(\partial_{(M}\cH_{N)L}-\half\partial_{L}\cH_{MN})+\epsilon^{ij}\partial_{i}\cA_{jL}\right]\,.
\ea
\label{forX}
\ee
Every line in (\ref{forEOM}) then corresponds to  \textbf{\textit{the equation of motion for each field}} as follows. 
\begin{itemize}
\item For the world-sheet metric, $h_{ij}$, we have the Virasoro constraints, 
\be
\left(\rdA_{i}X^{M}\rdA_{j}X^{N}-\half h_{ij}\rdA_{k}X^{M}\rdA^{k}X^{N}\right)\cH_{MN}=0\,.
\ee
\item For the gauge connection, $\cA_{iM}$,  since it  is not  arbitrary   but  derivative-index-valued,  from
\be
\ba{ll}
\cA_{i}^{M}\delta\cA_{jM}=0\,,\quad&\quad\delta\cA_{i}^{M}\partial_{M}\Phi=0\,,
\ea
\label{cAdcA}
\ee
the equation of motion amounts to 
\be
\delta\cA_{iM}\left(\cH^{M}{}_{N} \rdA^{i}X^{N}+\textstyle{\frac{1}{\sqrt{-h}}}\epsilon^{ij}\rdA_{j}X^{M}\right)=0\,.
\label{EOMA}
\ee
That is to say, the quantity inside the bracket should be  derivative-index-valued too. 
\item For the dynamical field, $X^{L}$, from (\ref{1HG}),
\be
\textstyle{\frac{1}{\sqrt{-h}}}\partial_{i}\left(\sqrt{-h}\rdA^{i}X^{M}\cH_{ML}\right)
-2\Gamma_{LMN}(P\rdA_{i}X)^{M}(\brP\rdA^{i}X)^{N}+
\textstyle{\frac{1}{\sqrt{-h}}}\epsilon^{ij}\partial_{i}\cA_{jL}=0\,,
\ee
or equivalently from (\ref{forX}), (\ref{3HG}),
\be
\!\!\!\!\!\!\!\!\!\textstyle{\frac{1}{\sqrt{-h}}}\partial_{i}(\sqrt{-h}\rdA^{i}X^{N})
+2\Gamma_{KLM}\!\left[(P\rdA_{i}X)^{K}\brP^{LN}+(\brP\rdA_{i}X)^{K}P^{LN}\right]\!\rdA^{i}X^{M}+\textstyle{\frac{1}{\sqrt{-h}}}\epsilon^{ij}\partial_{i}\cA_{jM}\cH^{MN}=0\,.
\ee
Here, with a pair of projectors, 
\be
\ba{ll}
P_{AB}=\half(\cJ+\cH)_{AB}\,,\quad&\quad
\brP_{AB}=\half(\cJ-\cH)_{AB}\,,
\ea
\ee
we set 
\be
\ba{ll}
(P\rdA_{i}X)^{M}=P^{M}{}_{N}\rdA_{i}X^{N}\,,\quad&\quad
(\brP\rdA_{i}X)^{M}=\brP^{M}{}_{N}\rdA_{i}X^{N}\,,
\ea
\ee
and  $\Gamma_{LMN}$ is the DFT analogy of  the Christoffel connection~\cite{Jeon:2011cn},\footnote{See Appendix \ref{AppendixCOVD} for a concise   review of the covariant derivatives in DFT.}
\be
\ba{ll}
\Gamma_{CAB}=&2\left(P\partial_{C}P\brP\right)_{[AB]}
+2\left({\brP_{[A}{}^{D}\brP_{B]}{}^{E}}-{P_{[A}{}^{D}P_{B]}{}^{E}}\right)\partial_{D}P_{EC}\\
{}&~-\textstyle{\frac{4}{D-1}}\left(\brP_{C[A}\brP_{B]}{}^{D}+P_{C[A}P_{B]}{}^{D}\right)\!\left(\partial_{D}d+(P\partial^{E}P\brP)_{[ED]}\right)\,.
\ea
\ee
\end{itemize}
~\\

\section{Reductions to undoubled formalisms\label{secReduction}}
In this section, we consider  reductions of the doubled formalism to undoubled formalisms by solving the section condition~(\ref{wcon}), (\ref{stcon})    explicitly:  we require  all the target spacetime fields, including the generalized metric, to be  independent of the dual coordinates,
\be
\frac{\,\partial\Phi(\ty,y)\,}{\partial \ty_{\mu}}=0\,.
\label{secf}
\ee
As a consequence,  the latter half of the components of the derivative-index-valued gauge connection is trivial,\footnote{Note   the ordering of the ordinary and the dual coordinates in our convention,   $x^{M}=(\ty_{\mu},y^{\nu})$, $\partial^{M}=\cJ^{MN}\partial_{N}=(\partial_{\mu},\tilde{\partial}^{\nu})$.}
\be
\cA^{M}=A_{\lambda}{}\partial^{M}y^{\lambda}=\left(A_{\mu}\,,\,0\right)\,,
\label{Aform}
\ee
and the gauged differential one-forms are explicitly (\textit{c.f.~}\cite{Giveon:1991jj,Hull:2006va}),
\be
\rdA_{i}X^{M}=\left(\partial_{i}\tY_{\mu}-A_{i\mu}\,,\,\partial_{i}Y^{\nu}\right)\,.
\label{rdAX}
\ee
Now we turn to the parametrization of the generalized metric which must  satisfy  the defining properties~(\ref{defcH}),
\be
\ba{lll}
\cH_{AB}=\left(\ba{cc}
U^{\mu\nu}&N^{\mu}{}_{\lambda}\\
(N^{t}){}_{\rho}{}^{\nu}&S_{\rho\lambda}
\ea
\right)\,,\quad&\quad\cH_{AB}=\cH_{BA}\,,
\quad&\quad\cH_{A}{}^{B}\cH_{B}{}^{C}=\delta_{A}{}^{C}\,.
\ea
\ee
With the  fixing of the section  by (\ref{secf}), (\ref{Aform}) and (\ref{rdAX}), depending on  whether the upper left $D\times D$ block, \textit{i.e.~}$U^{\mu\nu}$,   is degenerate or not,   there are two classes of parametrizations.  Each of them  requires  separate analysis.

\subsection{Non-degenerate case: Reduction  to  standard form}
As emphasized in \cite{Jeon:2010rw} and also   recently discussed  in  \cite{Godazgar:2013bja},  when  the upper left $D\times D$ block of the generalized metric, $U^{\mu\nu}$,  is non-degenerate, the remaining parts are  completely determined by the non-degenerate symmetric  matrix and  one free anti-symmetric matrix which we may identify with the usual  Riemannian metric and the Kalb-Ramond $B$-field respectively,   
\be
\ba{llll}
\cH_{AB}=\left(\ba{cc}
G^{-1}&-G^{-1}B\\
BG^{-1}&~G-BG^{-1}B
\ea
\right)\,,\quad&~~~ G_{\mu\nu}=G_{\nu\mu}\,,\quad&~~~ \det(G_{\mu\nu})\neq0\,,
\quad&~~~
B_{\mu\nu}=-B_{\nu\mu}\,.
\ea
\label{nondegH}
\ee
With this `standard' parametrization of the generalized metric,  the doubled yet gauged sigma model~(\ref{Lagrangian}) reduces to an expression which is multiplied by   the `correct' value of the string   tension, 
\be
\textstyle{\frac{1}{4\pi\alpha^{\prime}}}\cL
\equiv{\textstyle{\frac{1}{2\pi\alpha^{\prime}}}}\cL^{\prime}\,,
\label{tensionLprime}
\ee
\be
\ba{ll}
\cL^{\prime}=&-\half\sqrt{-h}h^{ij}
\partial_{i}Y^{\mu}\partial_{j}Y^{\nu}G_{\mu\nu}(Y)+\half\epsilon^{ij}\partial_{i}Y^{\mu}\partial_{j}Y^{\nu}B_{\mu\nu}(Y)+\half\epsilon^{ij}\partial_{i}\tY_{\mu}\partial_{j}Y^{\mu}\\
{}&
-\quarter\sqrt{-h}h^{ij}(\cC_{i\mu}-A_{i\mu})(\cC_{j\nu}-A_{j\nu})G^{\mu\nu}(Y)\,.
\ea
\label{Lprime}
\ee
Here $\cC_{i\mu}$ denotes the  on-shell value of the connection, 
\be
\cC_{i\mu}=\partial_{i}\tY_{\mu}+\partial_{i}Y^{\lambda}B_{\lambda\mu}+\textstyle{\frac{1}{\sqrt{-h}}}\epsilon_{i}{}^{j}\partial_{j}Y^{\lambda}G_{\lambda\mu}\,.
\label{cCcA}
\ee
It is also useful to note
\be
\rdA_{i}X^{M}\rdA_{j}X^{N}\cH_{MN}=\partial_{i}Y^{\mu}\partial_{j}Y^{\nu}G_{\mu\nu}+
(\rdA_{i}\tY_{\mu}+\partial_{i}Y^{\lambda}B_{\lambda\mu})
(\rdA_{j}\tY_{\nu}+\partial_{j}Y^{\rho}B_{\rho\nu})G^{\mu\nu}\,.
\ee
Integrating out the connection, we may ignore the second line in (\ref{Lprime}). The remaining  first line then consists of  the standard undoubled  string Lagrangian  without the dual coordinate fields, $\tY_{\mu}$, and a total derivative involving the dual fields, 
\be
\partial_{i}\left(\half\epsilon^{ij}\tY_{\mu}\partial_{j}Y^{\mu}\right)=\partial_{j}\left(\half\epsilon^{ij}Y^{\mu}\partial_{i}\tY_{\mu}\right)\,.
\label{topterm}
\ee
In fact, this total derivative   is precisely the topological term introduced  in \cite{Giveon:1991jj,Hull:2006va}  for the quantum equivalence of a doubled sigma model to the usual  formalism for world-sheets of arbitrary genus. Our covariant action naturally reproduces it from the first principle of the `coordinate gauge symmetry': The topological  term shares the same geometric origin as the world-sheet pull-back of the $B$-field. \\

Furthermore,  with $\rdA_{i}\tY_{\mu}=\partial_{i}\tY_{\mu}-A_{i\mu}$~(\ref{rdAX}), the on-shell value of the connection~(\ref{cCcA}) reads
\be
\rdA_{i}\tY_{\mu}+\partial_{i}Y^{\lambda}B_{\lambda\mu}+\textstyle{\frac{1}{\sqrt{-h}}}\epsilon_{i}{}^{j}\partial_{j}Y^{\lambda}G_{\lambda\mu}=0\,.
\label{SD0}
\ee
Since $G_{\mu\nu}$ is \textit{non-degenerate},  this relation implies
\be
G^{\mu\nu}\rdA_{i}\tY_{\nu}-(G^{-1}B)^{\mu}{}_{\nu}\partial_{i}Y^{\nu}
+\textstyle{\frac{1}{\sqrt{-h}}}\epsilon_{i}{}^{j}\partial_{j}Y^{\mu}=0\,,
\label{SD1}
\ee
and hence in particular, 
\be
(BG^{-1})_{\mu}{}^{\nu}\rdA_{i}\tY_{\nu}-(BG^{-1}B)_{\mu\nu}\partial_{i}Y^{\nu}
+\textstyle{\frac{1}{\sqrt{-h}}}\epsilon_{i}{}^{j}B_{\mu\nu}\partial_{j}Y^{\nu}=0\,.
\label{SD1p}
\ee
In addition to this result,  Eq.(\ref{SD0}) also implies, after contraction with $\textstyle{\frac{1}{\sqrt{-h}}}\epsilon_{k}{}^{i}$ and using  (\ref{epep}),
\be
\textstyle{\frac{1}{\sqrt{-h}}}\epsilon_{k}{}^{i}(\rdA_{i}\tY_{\mu}-B_{\mu\nu}\partial_{i}Y^{\nu})+G_{\mu\nu}\partial_{k}Y^{\nu}=0\,.
\label{SD1pp}
\ee
Finally, combining  (\ref{SD1p}) and (\ref{SD1pp}), we acquire
\be
(BG^{-1})_{\mu}{}^{\nu}\rdA_{i}\tY_{\nu}+(G-BG^{-1}B)_{\mu\nu}\partial_{i}Y^{\nu}+\textstyle{\frac{1}{\sqrt{-h}}}\epsilon_{i}{}^{j}\rdA_{j}\tY_{\mu}=0\,.
\label{SD2}
\ee
In terms of the doubled coordinates and the generalized metric, the two equations, (\ref{SD0}) and (\ref{SD2}),  exactly amount to  a self-duality relation in the doubled spacetime! \textit{c.f.~}(\ref{EOMA}),
\be
\cH^{M}{}_{N}\rdA_{i}X^{N}+\textstyle{\frac{1}{\sqrt{-h}}}\epsilon_{i}{}^{j}\rdA_{j}X^{M}=0\,.
\label{SD}
\ee
Thus,  for the \textit{non-degenerate cases}  the full set of self-duality relations~\eqref{SD}  follows from the equation of motion~(\ref{SD0}) without being  imposed   by hand. On the other hand  for \textit{degenerate cases},  as we see below and also  expected from the generic expression~(\ref{EOMA}),  this is not true in general: not all the components of the relation (\ref{SD}) are satisfied.     \\

\subsection{Degenerate case: Non-Riemannian geometry}
We start with an exact solution   of supergravity obtained  in \cite{Dabholkar:1990yf} which corresponds to  a macroscopic fundamental string geometry with the   $\SO(1,1) \times \SO(8)$ isometry. In string frame, the  background  reads
\be
\ba{l}
\rd s^{2} = f^{-1} \left(-\rd t^{2} + (\rd x^{1})^{2} \right) + (\rd x^{2})^{2} + \cdots + (\rd x^{9})^{2}\,,
\\
B =  (f^{-1} + c)\,\rd t\wedge\rd x^{1}\,,
\\
e^{-2\phi} = c^{\prime} f\,,
\ea\label{fsol}
\ee
where  $f$ is a harmonic function,
\be
\ba{ll}
f= 1+\frac{Q}{r^{6}}\,,\quad&\quad r^{2} = \sum_{a=2}^{9} (x^{a})^{2}\,,
\ea
\ee 
and $c$ and $c^{\prime}$ are arbitrary constants. In particular, the former corresponds to a `non-physical' constant shift of the $B$-field which  we introduce.   The corresponding generalized metric is then
\be
\cH_{MN} =\left(\ba{cccc}
f \eta^{\alpha\beta}&0&  - (1+c f) \cE^{\alpha}{}_{\delta}&0\\
0&\delta^{ab}&0&0\\
(1+ c f)\cE_{\gamma}{}^{\beta} &0 & - c (2 +c f) \eta_{\gamma\delta}&0\\
0&0&0&\delta_{cd}\ea\right)\,,
\label{F1geo}
\ee
and the DFT-dilaton is given by
\be
e^{-2d}=e^{-2\phi}\sqrt{-g}=c^{\prime}\,.
\ee
Here the greek letters,  $\alpha,\beta,\gamma,\delta, \cdots$, denote the  Minkowskian $\SO(1,1)$ vector indices subject to  the flat metric, $\eta_{\alpha\beta}=\mbox{diag}(-+)$, and the roman letters,  $a,b,c,d, \cdots$, are for the Euclidean   $\SO(8)$ vector indices with the   Kronecker-delta flat metric.  As seen in (\ref{F1geo}), the generalized metric  then decomposes into  sixteen blocks,  $(2+8+2+8)\times (2+8+2+8)$.  
 Further, with the $2\times 2$ anti-symmetric Levi-Civita symbol, $\cE_{\alpha\beta}=-\cE_{\beta\alpha}$,  $\cE_{01}=+1$, we set 
\be
\cE^{\alpha}{}_{\beta} = \eta^{\alpha\gamma}\cE_{\gamma \beta}=  -\cE_{\beta}{}^{\alpha}=-\cE_{\beta\delta}\eta^{\delta\alpha}\,,
\ee
which satisfies
\be
\cE^{\alpha}{}_{\beta}\cE^{\beta}{}_{\gamma}=\delta^{\alpha}{}_{\gamma}\,.
\ee
~\\

\noindent Now we perform an $\ODD$ rotation  exchanging  $(t,x^{1})$ and $(\tilde{t},\tx_{1})$ planes,\footnote{For discussion on T-duality along the temporal direction, see \textit{e.g.~}\cite{Moore:1993zc}.}
\be
\ba{ll}
\cH_{AB}\quad\longrightarrow\quad\cO_{A}{}^{C}\cO_{B}{}^{D}\cH_{CD}\,,\quad&\quad
\cO_{A}{}^{B}=\left(\ba{cccc}
0&0&\eta^{\mu\nu}&0\\
0&\delta^{a}{}_{b}&0&0\\
\eta_{\lambda\rho}&0&0&0\\
0&0&0&\delta_{c}{}^{d}
\ea
\right)\,,
\ea
\ee
and obtain a T-dual background,
\be
\cH_{MN} =\left(\ba{cccc}
-c(2+cf) \eta^{\alpha\beta}&0&   (1+c f) \cE^{\alpha}{}_{\delta}&0\\
0&\delta^{ab}&0&0\\
-(1+ c f)\cE_{\gamma}{}^{\beta} &0 & f\eta_{\delta\gamma}&0\\
0&0&0&\delta_{cd}\ea\right)\,,
\label{dualgeom}
\ee
which corresponds to
\be
\ba{ll}
\rd s^{2} = - \frac{1}{c(2+cf)}\left(-\rd t^{2} + (\rd x^{1})^{2} \right) + (\rd x^{2})^{2} + \cdots + (\rd x^{9})^{2}\,,
\\
B = \frac{1+ cf}{c(2+cf)}\,\rd t\wedge\rd x^{1}\,,
\\
e^{-2\phi} = c^{\prime} c(2+cf)\,.
\ea\label{dualgeom}
\ee
Clearly,  as long as  $c\neq 0$, the Riemannian metric is non-degenerate and well-defined. The corresponding doubled yet gauged  sigma model then can be readily read off from the  results above, (\ref{nondegH}), (\ref{Lprime}).  On the other hand,  when we take the limit,  $c\rightarrow 0$,   the Riemannian metric becomes   everywhere singular.\footnote{Nevertheless  we merely note  that the combination, $e^{-2\phi}g_{\mu\nu}$,  is  finite.  See also \cite{Malek:2013sp} for other examples of singular metrics.} But,   the corresponding generalized metric is perfectly smooth and  reduces in the limit  to  
\be
\cH_{MN} =\left(\ba{cccc}
0&0&    \cE^{\alpha}{}_{\delta}&0\\
0&\delta^{ab}&0&0\\
-\cE_{\gamma}{}^{\beta} &0 & f\eta_{\delta\gamma}&0\\
0&0&0&\delta_{cd}\ea\right)\,.
\label{dualgeom2}
\ee
A remarkable fact is then that, despite the  ``Riemannian" singularity,  the  doubled yet gauged  sigma model~(\ref{Lagrangian}) can  describe a string  propagating in such a  background well,   explicitly  through   
\be
\textstyle{\frac{1}{4\pi\alpha^{\prime}}}\cL
\equiv{\textstyle{\frac{1}{2\pi\alpha^{\prime}}}}\cL^{\prime\prime}\,,
\label{tensionLprimeprime}
\ee
\be
\ba{rl}
\cL^{\prime\prime}=&-\quarter\sqrt{-h}h^{ij}
\partial_{i}Y^{\alpha}\partial_{j}Y^{\beta}\eta_{\alpha\beta}f(Y)
-\half\sqrt{-h}h^{ij}\partial_{i}\tilde{Y}_{\alpha}\partial_{j}Y^{\beta}\cE^{\alpha}{}_{\beta}\\
{}&+\half\sqrt{-h}A_{i\alpha}\left(
\cE^{\alpha}{}_{\beta}h^{ij}\partial_{j}Y^{\beta}+\textstyle{\frac{1}{\sqrt{-h}}}\epsilon^{ij}\partial_{j}Y^{\alpha}\right)\\
{}&-\half\sqrt{-h}h^{ij}
\partial_{i}Y^{a}\partial_{j}Y_{a}+\half\epsilon^{ij}\partial_{i}\tY_{a}\partial_{j}Y^{a}\\
{}&-\quarter\sqrt{-h}h^{ij}\left(\partial_{i}\tY_{a}+\textstyle{\frac{1}{\sqrt{-h}}}\epsilon_{i}{}^{k}\partial_{k}Y_{a}-A_{ia}\right)\left(\partial_{j}\tY^{a}+\textstyle{\frac{1}{\sqrt{-h}}}\epsilon_{j}{}^{l}\partial_{l}Y^{a}-A_{j}{}^{a}\right)\\
=&-\quarter\sqrt{-h}h^{ij}
\partial_{i}Y^{\alpha}\partial_{j}Y^{\beta}\eta_{\alpha\beta}f(Y)
-\half\sqrt{-h}h^{ij}
\partial_{i}Y^{a}\partial_{j}Y_{a}+\half\epsilon^{ij}\partial_{i}\tY_{\mu}\partial_{j}Y^{\mu}
\\
{}&+\half\sqrt{-h}(A_{i\alpha}-\partial_{i}\tY_{\alpha})\left(
\cE^{\alpha}{}_{\beta}h^{ij}\partial_{j}Y^{\beta}+\textstyle{\frac{1}{\sqrt{-h}}}\epsilon^{ij}\partial_{j}Y^{\alpha}\right)\\
{}&-\quarter\sqrt{-h}h^{ij}\left(\partial_{i}\tY_{a}+\textstyle{\frac{1}{\sqrt{-h}}}\epsilon_{i}{}^{k}\partial_{k}Y_{a}-A_{ia}\right)\left(\partial_{j}\tY^{a}+\textstyle{\frac{1}{\sqrt{-h}}}\epsilon_{j}{}^{l}\partial_{l}Y^{a}-A_{j}{}^{a}\right)\,.
\ea
\label{Lprimeprime}
\ee
We note that, while the $\SO(8)$ sector of $\{Y^{a}, \tilde{Y}_{a}, A_{ia}\}$ agrees with the non-degenerate result~(\ref{Lprime}) having  the flat Euclidean metric, $\delta_{ab}$, and  the vanishing $B$-field, the $\SO(1,1)$ sector of $\{Y^{\alpha}, \tilde{Y}_{\alpha}, A_{i\alpha}\}$ takes a novel exotic form.  In particular, the gauge field components are  quadratic for the non-degenerate $\SO(8)$ sector, whereas  they are linear for  the degenerate $\SO(1,1)$ sector.  \\

\noindent Integrating out all the gauge fields, the doubled yet gauged  sigma model reduces to
\be
\textstyle{\frac{1}{2\pi\alpha^{\prime}}}\left[-\quarter\sqrt{-h}h^{ij}
\partial_{i}Y^{\alpha}\partial_{j}Y^{\beta}\eta_{\alpha\beta}f(Y)
-\half\sqrt{-h}h^{ij}
\partial_{i}Y^{a}\partial_{j}Y_{a}+\half\epsilon^{ij}\partial_{i}\tY_{\mu}\partial_{j}Y^{\mu}\right]\,,
\label{Ldeg}
\ee
where now the   two of the   `ordinary' coordinate fields  must satisfy   a self-duality  constraint,
\be
\partial_{i}Y^{\alpha}+\textstyle{\frac{1}{\sqrt{-h}}}\epsilon_{i}{}^{j}\cE^{\alpha}{}_{\beta}\partial_{j}Y^{\beta}=0\,.
\label{SDcE}
\ee
This is in contrast to  the  non-degenerate  $\SO(8)$ sector of which the  ordinary and the dual  coordinate fields are, like (\ref{SD0}), related by  a  different   type of a self-duality  relation, 
\be
\ba{lll}
D_{i}\tY_{a}+\textstyle{\frac{1}{\sqrt{-h}}}\epsilon_{i}{}^{j}\partial_{j}Y_{a}=0\quad&\Longleftrightarrow&\quad
\partial_{i}Y_{a}+\textstyle{\frac{1}{\sqrt{-h}}}\epsilon_{i}{}^{j}D_{j}\tY_{a}=0\,,
\ea
\ee
To summarize, even for the degenerate sector for which the Riemannian metric is ill-defined,   there exists a sigma model type  Lagrangian description  involving  a self-duality  constraint.  \\

\noindent In order to illustrate this feature in a more general setup, let us consider an extreme case of the  degeneracy where the upper left $D\times D$ block   of a generalized metric    vanishes itself,
\be
\ba{llll}
\cH_{AB} = \begin{pmatrix} 0 & N^{\mu}{}_{\lambda} \\ (N^{t})_{\rho}{}^{\nu} & S_{\rho\lambda} \end{pmatrix}\,,\quad&\quad N^2=1\,,\quad&\quad S=S^{t}\,,\quad&\quad SN=-(SN)^{t}\,.
\ea
\ee
In this background, the doubled yet gauged  sigma model reduces to 
\be
\ba{ll}
\textstyle{\frac{1}{4\pi\alpha^{\prime}}}\cL
\equiv{\textstyle{\frac{1}{2\pi\alpha^{\prime}}}}\Big[&
-\quarter\sqrt{-h}h^{ij}
\partial_{i}Y^{\mu}\partial_{j}Y^{\nu}S_{\mu\nu}(Y)
+\half\epsilon^{ij}\partial_{i}\tY_{\mu}\partial_{j}Y^{\mu}
\\
{}&
~~+\half\sqrt{-h}(A_{i\mu}-\partial_{i}\tY_{\mu})\left(
N^{\mu}{}_{\nu}h^{ij}\partial_{j}Y^{\nu}+\textstyle{\frac{1}{\sqrt{-h}}}\epsilon^{ij}\partial_{j}Y^{\mu}\right)~\Big]\,.
\ea
\label{LNS}
\ee
Integrating out the gauge field which is linear in the Lagrangian,  we end up with the first line which consists of the topological term and  a usual sigma model kinetic term for the ordinary coordinate fields with the halved tension. In addition, the ordinary coordinate fields must satisfy a self-dual relation,
\be
\partial_{i}Y^{\mu}+\textstyle{\frac{1}{\sqrt{-h}}}\epsilon_{i}{}^{j}N^{\mu}{}_{\nu}\partial_{j}Y^{\nu}=0\,.
\label{SDN}
\ee
These agree with the $\SO(1,1)$ sector of the above example, (\ref{Ldeg}), (\ref{SDcE}).\\

\noindent Especially for the trivial case of $S_{\rho\lambda}=0$,  the resulting reduced Lagrangian is purely topological,
\be
\textstyle{\frac{1}{4\pi\alpha^{\prime}}}\cL
\equiv{\textstyle{\frac{1}{4\pi\alpha^{\prime}}}}\epsilon^{ij}\partial_{i}\tY_{\mu}\partial_{j}Y^{\mu}\,.
\ee
~\\

\section{Summary and comments\label{secCON}}
We summarize the main results with  comments.
\begin{itemize}
\item  The coordinate gauge symmetry~(\ref{equivCGS}) is equivalent to the section condition, both (\ref{wcon}) and (\ref{stcon}).

\item The  coordinate gauge symmetry implies that \textit{\,spacetime is  doubled yet gauged.}

\item \textit{Gauged differential one-forms for the doubled coordinates,} $\rdA x^{M}=\rd x^{M}-\cA^{M}$~(\ref{GDO})  are fully covariant  under the generalized diffeomorphisms  and the coordinate gauge symmetry. In particular,  the gauge connection, $\cA^{M}$, is \textit{derivative-index-valued}~(\ref{divA}).

\item The completely covariant Lagrangian description of  a string  propagating   in the doubled yet gauged spacetime is given by~(\ref{Lagrangian}), \textit{i.e.}
\be
{\textstyle{\frac{1}{4\pi\alpha^{\prime}}}}\cL={\textstyle{\frac{1}{4\pi\alpha^{\prime}}}}\left[-\half\sqrt{-h}h^{ij}\rdA_{i}X^{M}\rdA_{j}X^{N}\cH_{MN}(X)-\epsilon^{ij}\rdA_{i}X^{M}\cA_{jM}
\right]\,.
\label{CCL}
\ee

\item For non-degenerate cases, the covariant Lagrangian reduces to (\ref{Lprime}),   \textit{i.e.}
\be
\ba{ll}
{\textstyle{\frac{1}{2\pi\alpha^{\prime}}}}\cL^{\prime}={\textstyle{\frac{1}{2\pi\alpha^{\prime}}}}\Big[\!\!&-\half\sqrt{-h}h^{ij}
\partial_{i}Y^{\mu}\partial_{j}Y^{\nu}G_{\mu\nu}(Y)+\half\epsilon^{ij}\partial_{i}Y^{\mu}\partial_{j}Y^{\nu}B_{\mu\nu}(Y)+\half\epsilon^{ij}\partial_{i}\tY_{\mu}\partial_{j}Y^{\mu}\\
{}&
-\quarter\sqrt{-h}h^{ij}(\cC_{i\mu}-A_{i\mu})(\cC_{j\nu}-A_{j\nu})G^{\mu\nu}(Y)~\Big]\,.
\ea
\label{LpB}
\ee
In particular, a self-duality condition~(\ref{SD}) relating the ordinary and the dual coordinate fields follows from the equations of motion,
\be
\rdA_{i}X^{M}+\textstyle{\frac{1}{\sqrt{-h}}}\epsilon_{i}{}^{j}\rdA_{j}X^{N}\cH_{N}{}^{M}=0\,.
\ee

\item Even for degenerate cases where the Riemannian metric becomes everywhere singular,  it is still possible to have a Lagrangian description of a string propagating in such backgrounds~(\ref{Lprimeprime}), (\ref{LNS}). Again a self-duality condition is implied by the equation of motion. However,  unlike the non-degenerate cases, it  is to be imposed only  on the ordinary coordinate fields, (\ref{SDcE}), (\ref{SDN}).

\item In both the non-degenerate and  the degenerate cases,  (\ref{topterm}),  \eqref{Ldeg},  \eqref{LNS},  there  appears   a  topological term  bi-linear in ordinary and dual  coordinates,  
\be
\textstyle{\frac{1}{4\pi\alpha^{\prime}}}\epsilon^{ij}\partial_{i}\tY_{\mu}\partial_{j}Y^{\mu}\,,
\ee
which was previously   introduced by hand   in \cite{Giveon:1991jj,Hull:2006va}. In our formalism, this term naturally arises  and   shares  the same geometric origin, \textit{i.e.~}the second term in (\ref{CCL}),  as the world-sheet pull-back of the $B$-field in (\ref{LpB}).

\item Especially for open string,  the topological term  leads to a  world-sheet boundary integral,
\be
-\textstyle{\frac{1}{4\pi\alpha^{\prime}}}\,\dis{\varointctrclockwise_{\partial\Sigma}\rd\sigma^{i}\,\tY_{\mu}\partial_{i}Y^{\mu}\,,}
\ee
which resembles the conventional  coupling of the string  to a (Born-Infeld)  gauge potential, $\varointctrclockwise\rd\sigma^{i}\,\hat{A}_{\mu}\partial_{i}Y^{\mu}$, and hints at an intriguing relation between  the dual coordinate and the  (Born-Infeld)  gauge potential,
\be
\ba{lll}
\left.\tY_{\mu}\right|_{\partial\Sigma}~&\Longleftrightarrow&~-4\pi\alpha^{\prime}\hat{A}_{\mu}\,.
\ea
\label{identify}
\ee 
In fact,  for a derivative-index-valued vector,  $\cV^{M}=\phi\partial^{M}\varphi$ (\ref{divV}), which generates  the coordinate gauge symmetry, the generalized Lie derivative~(\ref{tcL}) of  the  generalized metric, $\cH_{AB}$ (\ref{nondegH}), implies   (see ({2.9}) of   \cite{Park:2013mpa}), 
\be
\ba{llll}
\delta G_{\mu\nu}=0\,,&\quad\delta B_{\mu\nu}=\partial_{\mu}(\phi\partial_{\nu}\varphi)-\partial_{\nu}(\phi\partial_{\mu}\varphi)\,,&\quad\delta\tY_{\mu}=\phi\partial_{\mu}\varphi\,,&\quad \delta(B_{\mu\nu}-2\partial_{[\mu}\tY_{\nu]})=0\,,
\ea
\ee
of which the last would be, assuming  the identification   (\ref{identify}),  consistent with the  gauge invariant combination of the $B$-field and  the field strength, $B+4\pi\alpha^{\prime}\hat{F}$.

\item Our results are classical.  Quantization  remains as a future work,  especially for the degenerate non-Riemannian backgrounds.   

\item  The beta-functional world-sheet derivation of the DFT equations of motion, one-loop~\cite{Berman:2007xn,Copland:2011yh,Copland:2011wx} and beyond for the higher order $\alpha^{\prime}$-corrections~\cite{Hohm:2013jaa,Godazgar:2013bja},   may be now worth while to revisit   equipped  with the  full covariance.

\item Thorough investigation  of the degenerate geometry is desirable  within the frameworks of     both  the doubled yet gauged  sigma model and   the maximally   supersymmetric double field theory~\cite{Jeon:2012hp}.   It seems natural to expect    such a background  to  provide  an alternative or    enriched scheme for compactification.

\item Understanding of the coordinate gauge symmetry from the Hamiltonian view point  for a constrained system  deserves a separate study~\cite{BMR2013}.

\item Supersymmetrization as well as generalization to a generic $p$-brane  are also of interest,  \textit{e.g.}~using the methods of \cite{Rocek:1991ps,Hull:2006va,Park:2008qe}.
\end{itemize}
\section*{Acknowledgements} 
We wish to thank   David Berman, Chris Blair, Martin Cederwall,  Imtak Jeon, Sung-Sik Lee,   Emanuel Malek,  Diego Marques, Malcolm Perry and Alasdair Routh for helpful   discussions.    We  should like to also acknowledge our indebtness to Jim Stasheff for his careful proof reading of our second version of the manuscript with helpful  encouraging   comments.   JHP deeply   acknowledges the    hospitality   from  Hugh Osborn at  DAMTP, Cambridge  during  his sabbatical visit.  The work was supported by the National Research Foundation of Korea (NRF) grant funded by the Korea government (MSIP) with the Grant  No.  2005-0049409 (CQUeST),  No. 2012R1A2A2A02046739,  No. 2013R1A1A1A05005747 and No. 2012R1A6A3A03040350.



\newpage

\appendix
\begin{center}
{\textbf{\Large{\underline{Appendices A \& B}}}}
\end{center}

\section{Covariant derivatives  in DFT\label{AppendixCOVD}}
Here we review the covariant derivatives in DFT.\\

\noindent With a pair of symmetric projectors,
\be
\ba{ll}
P_{AB}=P_{BA}=\half(\cJ+\cH)_{AB}\,,\quad&\quad\brP_{AB}=\brP_{BA}=\half(\cJ-\cH)_{AB}\,,
\ea
\label{projection}
\ee
satisfying
\be
\ba{lll}
P_{A}{}^{B}P_{B}{}^{C}=P_{A}{}^{C}\,,
\quad&\quad\brP_{A}{}^{B}\brP_{B}{}^{C}=\brP_{A}{}^{C}\,,
\quad&\quad P_{A}{}^{B}\brP_{B}{}^{C}=0\,,
\ea
\ee
the DFT analogy of  the Christoffel connection reads~\cite{Jeon:2011cn}
\be
\ba{ll}
\Gamma_{CAB}=&2\left(P\partial_{C}P\brP\right)_{[AB]}
+2\left({\brP_{[A}{}^{D}\brP_{B]}{}^{E}}-{P_{[A}{}^{D}P_{B]}{}^{E}}\right)\partial_{D}P_{EC}\\
{}&~-\textstyle{\frac{4}{D-1}}\left(\brP_{C[A}\brP_{B]}{}^{D}+P_{C[A}P_{B]}{}^{D}\right)\!\left(\partial_{D}d+(P\partial^{E}P\brP)_{[ED]}\right)\,,
\ea
\label{Gamma}
\ee
which  defines the \textit{semi-covariant derivative}~\cite{Jeon:2010rw,Jeon:2011cn},
\be
\na_{C}T_{A_{1}A_{2}\cdots A_{n}}
:=\partial_{C}\Tw_{A_{1}A_{2}\cdots A_{n}}-\omega\Gamma^{B}{}_{BC}T_{A_{1}A_{2}\cdots A_{n}}+
\sum_{i=1}^{n}\,\Gamma_{CA_{i}}{}^{B}T_{A_{1}\cdots A_{i-1}BA_{i+1}\cdots A_{n}}\,.
\label{semi-covD}
\ee
The connection is the unique solution to the following requirements.\begin{itemize}
\item The semi-covariant derivative is compatible with the $\ODD$ metric,
\be
\ba{lll}
\nabla_{A}\cJ_{BC}=0~&\Longleftrightarrow&~\Gamma_{CAB}+\Gamma_{CBA}=0\,.
\ea
\label{nacJ}
\ee
\item  The semi-covariant derivative annihilates the whole NS-NS sector, \textit{i.e.}  the DFT-dilaton\footnote{Since $e^{-2d}$ is a scalar density with weight one, $\na_{A}d=-\half e^{2d}\nabla_{A}e^{-2d}=\partial_{A}d+\half \Gamma^{B}{}_{BA}$.} and the pair of projectors~(\ref{projection}),
 \be
\ba{lll}
\nabla_{A}d=0\,,\quad&\quad
\nabla_{A}P_{BC}=0\,,\quad&\quad\nabla_{A}\brP_{BC}=0\,.
\ea
\label{naNSNS}
\ee
\item The cyclic sum of the connection vanishes,
\be
\Gamma_{ABC}+\Gamma_{CAB}+\Gamma_{BCA}=0\,.
\label{cyclic}
\ee
\item Lastly, the connection belongs to the  kernels  of   rank-six projectors,
\be
\ba{ll}
\cP_{CAB}{}^{DEF}\Gamma_{DEF}=0\,,\quad&\quad \bcP_{CAB}{}^{DEF}\Gamma_{DEF}=0\,,
\ea
\label{kernel}
\ee
where 
\be
\ba{ll}
\cP_{CAB}{}^{DEF}=P_{C}{}^{D}P_{[A}{}^{[E}P_{B]}{}^{F]}+\textstyle{\frac{2}{D-1}}P_{C[A}P_{B]}{}^{[E}P^{F]D}\,,\\
\bcP_{CAB}{}^{DEF}=\brP_{C}{}^{D}\brP_{[A}{}^{[E}\brP_{B]}{}^{F]}+\textstyle{\frac{2}{D-1}}\brP_{C[A}\brP_{B]}{}^{[E}\brP^{F]D}\,.
\ea
\label{P6}
\ee
\end{itemize}
In particular, the two symmetric properties, (\ref{nacJ}) and (\ref{cyclic}),   enable us to replace the ordinary derivatives in the definition of the generalized Lie derivative (\ref{tcL}) by the semi-covariant derivatives~(\ref{semi-covD}),
\be
\hcL_{V}T_{A_{1}\cdots A_{n}}=V^{B}\na_{B}T_{A_{1}\cdots A_{n}}+\omega\na_{B}V^{B}T_{A_{1}\cdots A_{n}} +\sum_{i=1}^{n}(\na_{A_{i}}V_{B}-\na_{B}V_{A_{i}})T_{A_{1}\cdots A_{i-1}}{}^{B}{}_{A_{i+1}\cdots  A_{n}}\,.
\ee
The rank-six projectors satisfy the projection property, 
\be
\ba{ll}
{\cP_{CAB}{}^{DEF}\cP_{DEF}{}^{GHI}=\cP_{CAB}{}^{GHI}\,,}\quad&\quad{\bcP_{CAB}{}^{DEF}\bcP_{DEF}{}^{GHI}=\bcP_{CAB}{}^{GHI}\,.}
\ea
\ee
They  are also  symmetric and traceless,
\be
\ba{ll}
{\cP_{CABDEF}=\cP_{DEFCAB}=\cP_{C[AB]D[EF]}\,,}~~&~~{\bcP_{CABDEF}=\bcP_{DEFCAB}=\bcP_{C[AB]D[EF]}\,,} \\
{\cP^{A}{}_{ABDEF}=0\,,~~~~\,P^{AB}\cP_{ABCDEF}=0\,,}~~&~~
{\bcP^{A}{}_{ABDEF}=0\,,~~~~\,\brP^{AB}\bcP_{ABCDEF}=0\,.}
\ea
\label{symP6}
\ee
~\\
Now, under the infinitesimal DFT-coordinate   transformation set by the generalized Lie derivative, the semi-covariant derivative transforms as
\be
\dis{\delta(\na_{C}T_{A_{1}\cdots A_{n}})=
\hcL_{V}(\na_{C}T_{A_{1}\cdots A_{n}})+
\sum_{i=1}^{n}\,2(\cP{+\bcP})_{CA_{i}}{}^{BDEF}
\partial_{D}\partial_{[E}V_{F]}T_{A_{1}\cdots A_{i-1} BA_{i+1}\cdots A_{n}}\,.}
\label{anomalous}
\ee
The sum on the right hand side corresponds to a potentially  anomalous part against  the full covariance. Hence, in general, the semi-covariant derivative is  not necessarily  covariant.\footnote{However, (\ref{nacJ}) and (\ref{naNSNS}) are exceptions as the anomalous terms vanish identically,  thanks to (\ref{symP6}). }   However, since  the anomalous terms are  projected by the rank-six projectors which satisfy the properties in (\ref{symP6}),  it is  in fact  possible to eliminate  them.  Combined with the projectors,   the semi-covariant derivative  ---as the name indicates---  can be converted into   various fully covariant derivatives~\cite{Jeon:2011cn}:
\be
\ba{l}
P_{C}{}^{D}\brP_{A_{1}}{}^{B_{1}}\brP_{A_{2}}{}^{B_{2}}\cdots\brP_{A_{n}}{}^{B_{n}}
\DO_{D}T_{B_{1}B_{2}\cdots B_{n}}\,,\\
\brP_{C}{}^{D}P_{A_{1}}{}^{B_{1}}P_{A_{2}}{}^{B_{2}}\cdots P_{A_{n}}{}^{B_{n}}
\DO_{D}T_{B_{1}B_{2}\cdots B_{n}}\,,\\
P^{AB}\brP_{C_{1}}{}^{D_{1}}\brP_{C_{2}}{}^{D_{2}}\cdots\brP_{C_{n}}{}^{D_{n}}\DO_{A}T_{BD_{1}D_{2}\cdots D_{n}}\,,\\
\brP^{AB}{P}_{C_{1}}{}^{D_{1}}{P}_{C_{2}}{}^{D_{2}}\cdots{P}_{C_{n}}{}^{D_{n}}\DO_{A}T_{BD_{1}D_{2}\cdots D_{n}}\,,\\
P^{AB}\brP_{C_{1}}{}^{D_{1}}\brP_{C_{2}}{}^{D_{2}}\cdots\brP_{C_{n}}{}^{D_{n}}
\DO_{A}\DO_{B}T_{D_{1}D_{2}\cdots D_{n}}\,,\\
\brP^{AB}P_{C_{1}}{}^{D_{1}}P_{C_{2}}{}^{D_{2}}\cdots P_{C_{n}}{}^{D_{n}}
\DO_{A}\DO_{B}T_{D_{1}D_{2}\cdots D_{n}}\,.
\ea
\label{covT}
\ee
~\\

\section{Useful formulae}
For the DFT generalized diffeomorphism connection~(\ref{Gamma}), we have
\be
\ba{l}
\Gamma_{LJK}P^{J}{}_{M}\brP^{K}{}_{N}=(P\partial_{L}P\brP)_{MN}\,,\\
\Gamma_{LJK}\brP^{J}{}_{M}P^{K}{}_{N}=-(\brP\partial_{L}P P)_{MN}\,,\\
\partial_{L}\cH_{MN}=4\Gamma_{LJK}P^{J}{}_{(M}\brP^{K}{}_{N)}\,,\\
\cH_{L}{}^{K}\partial_{K}\cH_{MN}=4\left(
\Gamma_{IJK}P^{I}{}_{L}P^{J}{}_{(M}\brP^{K}{}_{N)}+\Gamma_{IJK}\brP^{I}{}_{L}\brP^{J}{}_{(M}P^{K}{}_{N)}\right)\,,
\ea
\label{1HG}
\ee
such that
\be
\ba{ll}
\cH_{M}{}^{K}\partial_{N}\cH_{KL}+\cH_{N}{}^{K}\partial_{M}\cH_{KL}&=-\cH_{LK}\partial_{N}\cH^{K}{}_{M}-\cH_{LK}\partial_{M}\cH^{K}{}_{N}\\\
{}&=4\left(\Gamma_{MJK}P^{J}{}_{[N}\brP^{K}{}_{L]}+\Gamma_{NJK}P^{J}{}_{[M}\brP^{K}{}_{L]}\right)\,,
\ea
\ee
and 
\be
\cH_{L}{}^{K}\partial_{K}\cH_{MN}+\cH_{M}{}^{K}\partial_{N}\cH_{KL}+\cH_{N}{}^{K}\partial_{M}\cH_{KL}=-4\left(\Gamma_{JK(M}P^{J}{}_{N)}\brP^{K}{}_{L}+
\Gamma_{JK(M}\brP^{J}{}_{N)}P^{K}{}_{L}\right)\,.
\label{3HG}
\ee
It is  also worth while to note
\be
\ba{l}
(PU)^{M}\na_{M}(\brP V)^{N}+(\brP U)^{M}\na_{M}(PV)^{N}\\
=\half U^{M}\left[
\partial_{M}V^{N}+2P_{M}{}^{K}\Gamma_{K}{}^{N}{}_{L}(\brP V)^{L}+2\brP_{M}{}^{K}\Gamma_{K}{}^{N}{}_{L}(P V)^{L}\right]-
\half (\cH U)^{M}\partial_{M}(\cH V)^{N}\,.
\ea
\ee
~\\
On the string world-sheet we have
\be
\ba{ll}
\epsilon_{i}{}^{j}\epsilon_{k}{}^{l}=(-h)(\delta_{i}{}^{l}\delta_{k}{}^{j}-
h_{ik}h^{jl})\,,\quad&\quad
\epsilon_{ij}\epsilon^{kl}=(-h)(\delta_{i}{}^{l}\delta_{j}{}^{k}-\delta_{i}{}^{k}\delta_{j}{}^{l})\,.
\ea
\ee
In particular,
\be
\left(\textstyle{\frac{1}{\sqrt{-h}}}\epsilon_{i}{}^{j}\right)\left(
\textstyle{\frac{1}{\sqrt{-h}}}\epsilon_{j}{}^{k}\right)=\delta_{i}{}^{k}\,.
\label{epep}
\ee


\end{document}